\documentclass[a4paper,12pt]{article}

\usepackage{amsmath}
\usepackage{epsfig}

\newcommand{\Pgg}{\ensuremath{\mathrm{\gamma}}}
\newcommand{\Pep}{\ensuremath{\mathrm{e}^+}}
\newcommand{\Pe}{\ensuremath{\mathrm{e}}}
\newcommand{\Pgpp}{\ensuremath{\mathrm{\pi^+}}}
\newcommand{\Pgmp}{\ensuremath{\mathrm{\mu^+}}}
\newcommand{\Pgm}{\ensuremath{\mathrm{\mu}}}
\newcommand{\Px}{\ensuremath{\mathrm{x}}}
\newcommand{\PX}{\ensuremath{\mathrm{X}}}

\begin{document}


\begin{center}
  {\Large\bf Search for Exotic Muon Decays\footnote{supported by the
      BMBF (06 TU 886), DFG (Mu 705/3, Graduiertenkolleg) and the UK
      Engineering and Physical Sciences Research Council.}}\\

  \bigskip\bigskip

  R.~Bilger$^1$, K.~F\"ohl$^2$, H.~Clement$^1$, M.~Cr\"oni$^1$,
  A.~Erhardt$^1$, R.~Meier$^1$, J.~P\"atzold$^1$, and
  G.J.~Wagner$^1$\\ 

  \bigskip

  $^1${\it Physikalisches Institut der Universit\"at T\"ubingen, Auf der
  Morgenstelle 14, D-72076 T\"ubingen, Germany}\\

  \medskip

  $^2${\it Department of Physics and Astronomy, University of Edinburgh,
  James Clerk Maxwell Building, The King's Buildings, Mayfield Road,
  Edinburgh, EH9 3JZ, United Kingdom}\\

\end{center}

\noindent {\bf Abstract:} Recently, it has been proposed that the
observed anomaly in the time distribution of neutrino induced
reactions, reported by the KARMEN collaboration, can be interpreted as
a signal from an exotic muon decay branch $\Pgmp\to\Pep\PX$. It has
been shown that this hypothesis gives an acceptable fit to the KARMEN
data if the boson \PX\ has a mass of $\rm m_\PX=103.9\,MeV/c^2$, close
to the kinematical limit. We have performed a search for the \PX\ 
particle by studying for the first time the very low energy part of
the Michel spectrum in \Pgmp\ decays.  Using a HPGe detector setup at
the $\mu{}E4$ beamline at PSI we find branching ratios $\rm
BR(\Pgmp\to\Pep\PX)<5.7\cdot10^{-4}$ (90\% C.L.)  for most of the
region $\rm 103\,MeV/c^2<m_\PX<105\,MeV/c^2$.

\bigskip


\bigskip


\vfil

\noindent Corresponding author:\\
          Ralph Bilger\\
          Tel: +49 7071 2976304, Fax:+49 7071 295373\\
          e-mail: ralph.bilger@uni-tuebingen.de

\newpage

\section{Introduction}
At the Rutherford Appleton Laboratory (RAL) the KARMEN collaboration
is studying neutrino-nuclear reactions, induced from the decay
products of positive pions, which are produced and stopped in the
proton beam dump. In 1995 KARMEN for the first time
reported~\cite{armbruster:1995} an anomaly in the time distribution of
single prong events concerning the time interval corresponding to muon
decay. Even with a much improved active detector shielding the anomaly
has persisted in new KARMEN data~\cite{zeitnitz:1998}.

This anomaly has been suggested to originate from the observation of a
hitherto unknown weakly interacting neutral and massive fermion,
called \Px, from a rare pion decay process $\Pgpp\to\Pgmp\Px$. After a
mean flight path of $\rm 17.5\,m$ \Px\ is registered in the KARMEN
calorimeter after $\rm t_{TOF}=(3.60\pm0.25)\,\mu{}s$ beam on target
by its decay resulting in visible energies of typically $\rm
T_{vis}=11-35\,MeV$. The observed velocity and the two-body kinematics
of the assumed pion decay branch lead to a mass $\rm
m_\Px=33.9\,MeV/c^2$, extremely close to the kinematical limit.

The hypothetical decay $\Pgpp\to\Pgmp\Px$ has been searched for at PSI
in a series of experiments using magnetic spectrometers by studying
muons from pion decay in flight~\cite{bilger:1995:plb, daum:1995,
  daum:1998}, the latest measurement resulting in an upper limit for
the branching ratio of $\rm BR(\Pgpp\to\Pgmp\Px)<1.2\cdot10^{-8}$
(95\% C.L.) \cite{daum:1998}. Combined with theoretical constraints
which assume no new weak interaction~\cite{barger:1995:plb:352} this
result rules out the existence of this rare pion decay branch if \Px\ 
is an isodoublet neutrino. However, if \Px\ is mainly isosinglet
(sterile), the branching ratio can be considerably
lower~\cite{barger:1995:plb:356}. From the number of observed \Px\ 
events in comparison with the total number of \Pgpp\ decays the KARMEN
collaboration gives a lower limit for the branching ratio of
$10^{-16}$.

Very recently Gninenko and Krasnikov have
proposed~\cite{gninenko:1998} that the observed time anomaly can also
be explained by an exotic \Pgm\ decay branch $\Pgmp\to\Pep\PX$
resulting in the production of a new, weakly interacting neutral boson
with mass $\rm m_\PX=103.9\,MeV/c^2$. They show that a second
exponential in the KARMEN time distribution with time constant equal
to the muon lifetime and shifted by the flight time of the
\PX-particle $\rm t_{TOF}=3.60\,\mu{}s$ gives an acceptable fit to the
KARMEN data. Considering three possible \PX-boson phenomenologies,
they predict branching ratios for $\Pgmp\to\Pep\PX$ in the order of
$10^{-2}$, if \PX\ is a scalar particle; $10^{-5}$, if \PX\ decays via
a hypothetical virtual charged lepton; and $10^{-13}$, if \PX\ decays
via two additional hypothetical neutral scalar bosons.

In this paper we present a direct experimental search for the \PX\ 
particle by studying the low energy part of the Michel spectrum
looking for a peak from mono-energetic positrons with energy $\rm
T_\Pe=(m_\Pgm^2+m_\Pe^2-m_\PX^2)/(2m_\Pgm)-m_\Pe=1.23\,MeV$
resulting from the two-body decay $\Pgmp\to\Pep\PX$.

In the past, searches for exotic two-body \Pgm\ decay modes have
already been performed~\cite{bryman:1986:prl} motivated by predictions
about the existence of light, weakly interacting bosons like axions,
majorons, Higgs particles, familons and Goldstone bosons resulting in
upper limits for the branching ratio of approximatley $3\cdot10^{-4}$
(90\% C.L.). However, these searches are not sensitive to the
suggested \PX\ boson with $\rm m_\PX=103.9\,MeV/c^2$ since the lowest
positron energy region studied was between 1.6 and 6.8\,MeV,
corresponding to the \PX\ mass region 103.5 to $\rm 98.3\,MeV/c^2$.

\section{The Experiment}

The basic idea is to stop a $\mu^+$ beam inside a germanium detector.
The low energy decay positrons of interest also deposit their entire
kinetic energy in the detector volume.  For a sizeable fraction of
events the subsequent annihilation radiation does not interact with
the detector thus preserving the positron energy information.

\begin{figure}
  \centerline{\epsfig{file=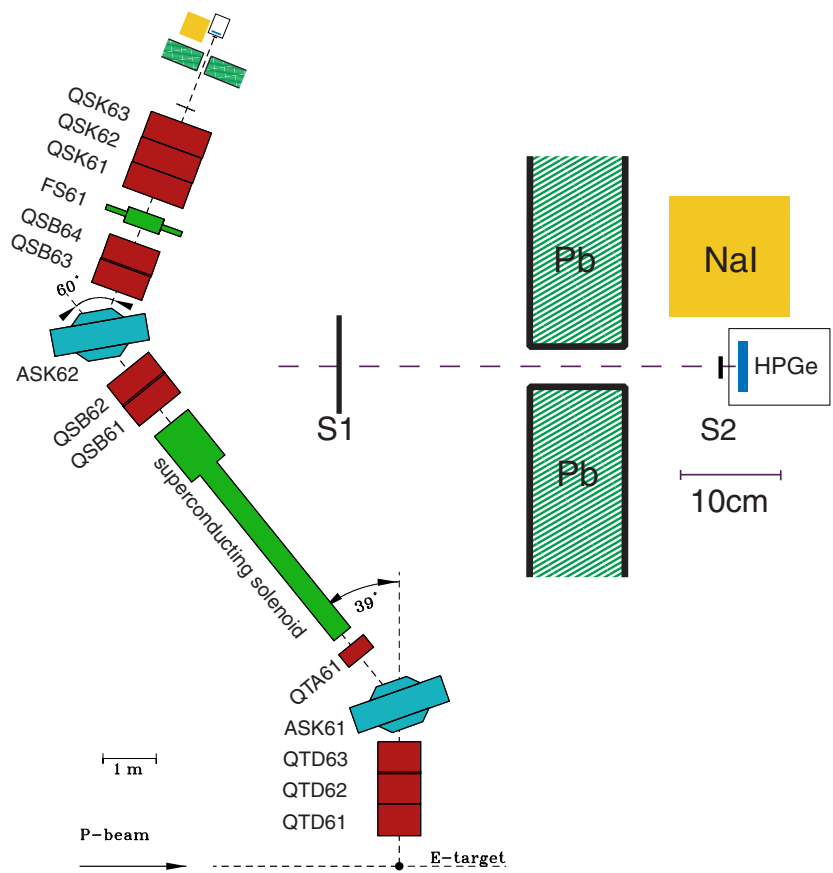,width=0.80 \textwidth}}
  \caption[]{Schematical layout of the experimental setup. The $\rm
    \Pgm{}E4$ low energy \Pgm\ channel at PSI is shown in the left
    part of the figure together with a sketch of the detector setup,
    which is shown in more detail in the right part of the figure.}
  \label{fig:setup}
\end{figure}

This experiment has been performed at the $\rm \mu{}E4$ channel at PSI
(see Fig.~\ref{fig:setup}). The beam line is optimized for intense
polarized muon beams in the momentum range between 30 and 100\,MeV/c
with very low pion and positron contamination. Pions from the
production target are collected at an angle of $90^\circ$ relative to
the primary proton beam and are injected into a long 5\,T
superconducting solenoid in which they can decay. The last part of the
beam line is the muon extraction section which allows the selection of
a central muon momentum different from that of the injected pions.

The detector setup consists of a large ($\rm 120\times200\,mm^2$)
2\,mm thick plastic scintillator counter S1 followed by a 35\,mm
diameter hole in a 10\,cm thick lead shielding wall and a small ($\rm
20\times20\,mm^2$) 1\,mm thick plastic scintillator counter S2
directly in front of a 9\,mm thick planar high purity germanium (HPGe)
detector with an area of $\rm 1900\,mm^2$. In addition, we have placed
a 127\,mm (5 inch) diameter, 127\,mm thick NaI detector shielded
against the \Pgm-flux adjacent to the HPGe for detecting 511\,keV
$\gamma$ rays from positron annihilation.

The coincidence $\rm S1\times{}S2\times{}HPGe$ was used as a trigger
which generated --- in addition to a prompt gate --- a delayed gate
$\rm 2.2-7.2\,\mu{}s$ after the prompt muon signal for the expected
decay events.  During the time period for the delayed gate, S1 was
used as a veto detector to discriminate against further beam
particles. Timing and energy information from the dectetors utilizing
several different methods for signal discrimination, amplification,
shaping and digitization were recorded for both prompt and delayed
signals using the MIDAS data acquisition system~\cite{midas:1998}.

For the energy calibration of signals occuring during the prompt gate,
\Pgg\ rays from $\rm ^{22}Na$ and $\rm ^{60}Co$ sources were used. In
order to derive the energy information from the HPGe detector signal,
both spectroscopy amplifiers and peak-sensitive ADCs as well as a
timing filter amplifier (TFA) connected to a charge sensitive QDC were
employed.  In addition, sample signals from the HPGe detector, both
before and after amplification, were recorded and stored with a
digital oscilloscope.  It turned out that every spectroscopy amplifier
available during the course of the experiment showed a significantly
varying baseline shift for a few microseconds following a prompt
signal. The variations of the baseline level just after the prompt
signal were due to fluctuations in time for the onset of the baseline
restoration circuitry.  Thus, for spectroscopy amplifiers, a
sufficiently accurate energy calibration for the delayed signal was
not possible. 

The TFA branch did not have such baseline problems, however the energy
resolution for the delayed signal in this branch is 100\,keV FWHM
only.  A short shaping time of $\rm 0.25\,\mu$s and low amplification
to avoid saturation from the high-amplitude prompt signal had to be
used to be ready in time for the delayed pulse.

During 12 hours of data taking $1.3\cdot10^7$ events were recorded on
tape. Saturation of the HPGe pre-amplifier at a singles rate of
$5-6\cdot10^3\,\rm s^{-1}$ was limiting the event rate.


\section{Results}

\begin{figure}
  \begin{center}
    \epsfig{file=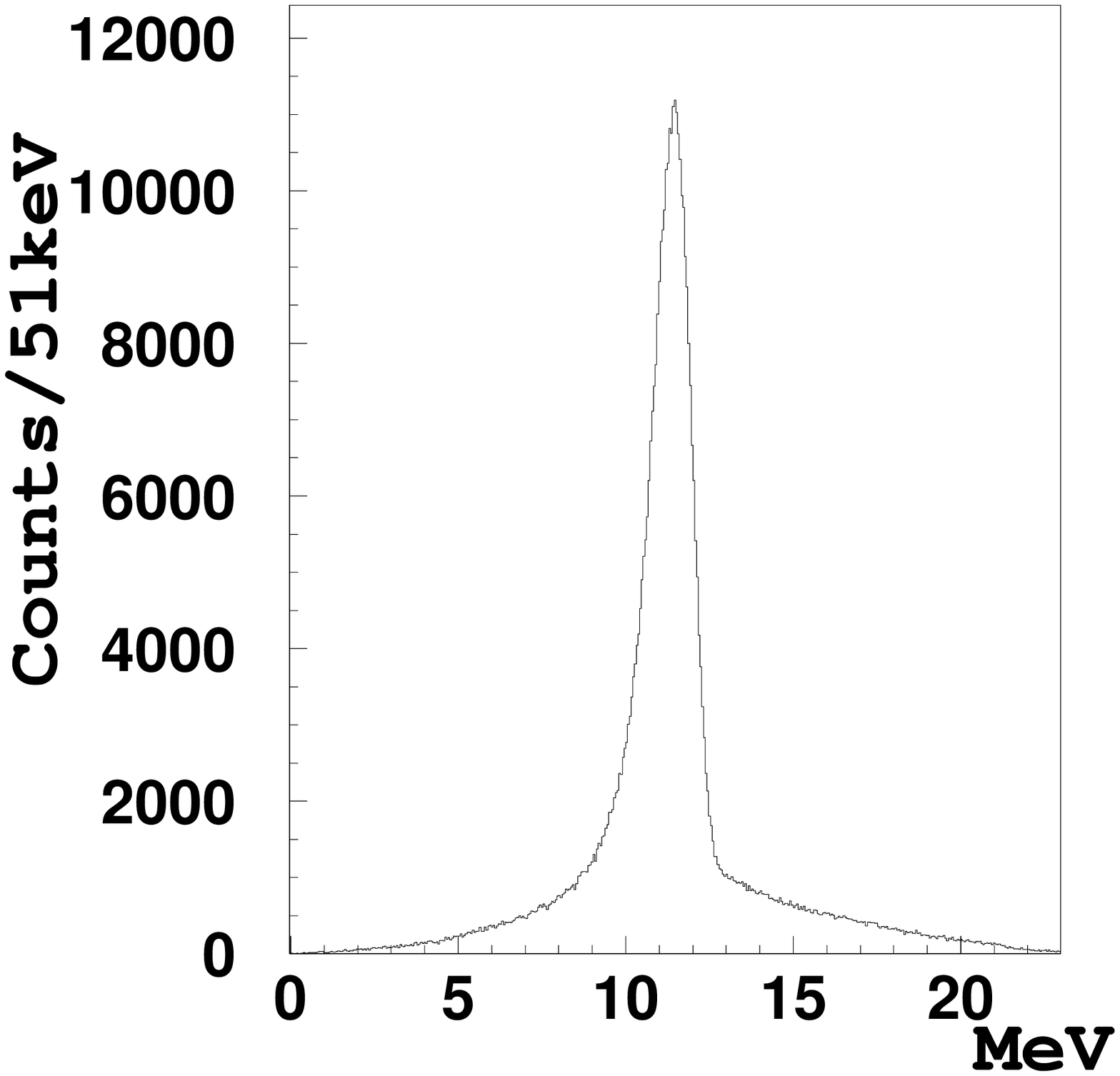,width=0.45 \textwidth,bbllx=32,bblly=160,bburx=560,bbury=648}
    \epsfig{file=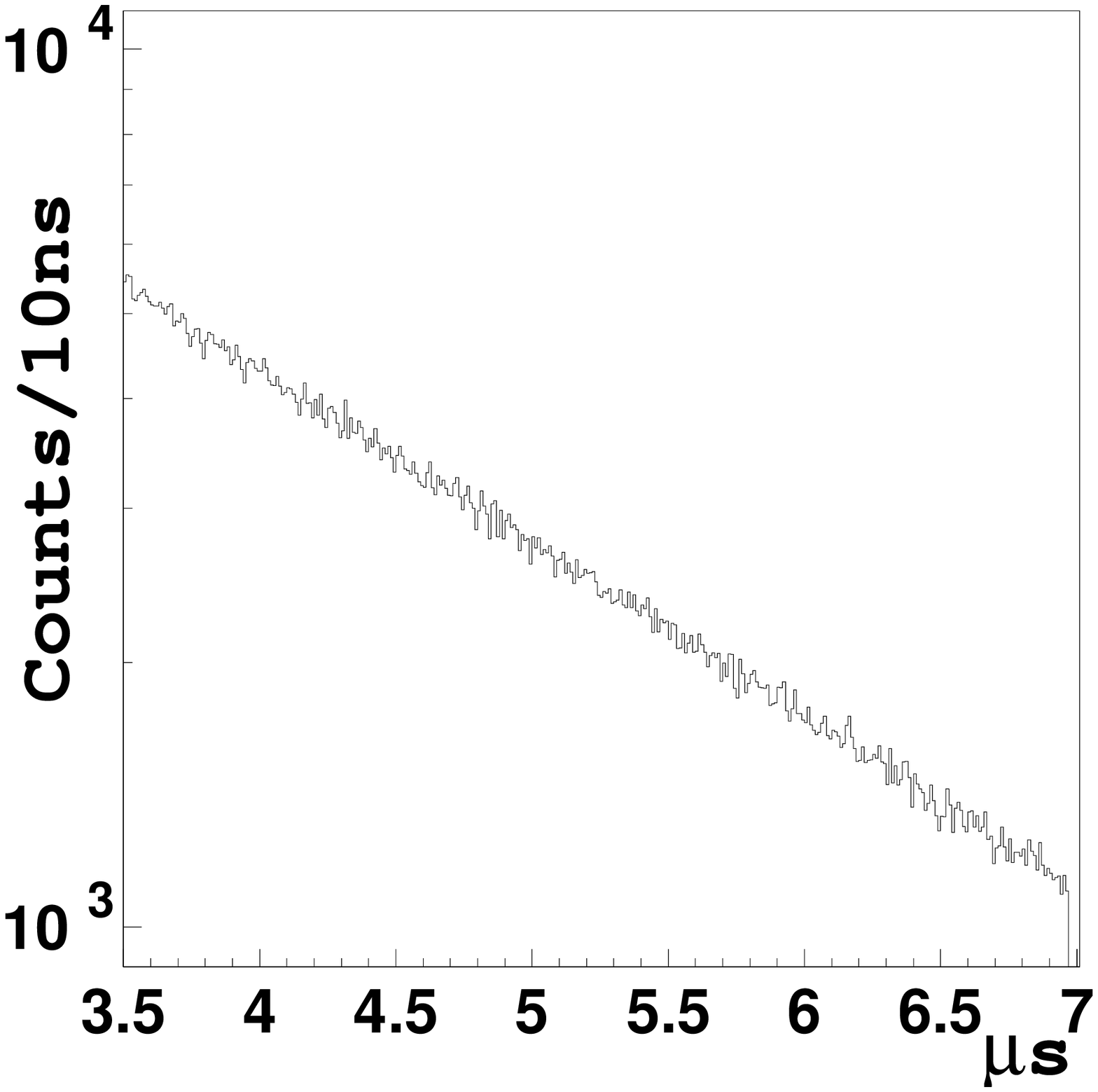,width=0.45 \textwidth,bbllx=0,bblly=160,bburx=528,bbury=648}
  \end{center}
  \caption[]{Left: Energy spectrum of prompt signals resulting from
    muons stopping in the HPGe detector with an additional constraint
    requiring the presence of an afterpulse arriving during the
    delayed gate.  Right: Spectrum for the time difference between
    delayed and prompt signals. The time constant $\tau=2.21\pm0.02\mu
    s$ of the exponential shape is in very good agreement with the
    muon life time.}
  \label{fig:muex_hpge_delayed_tdc_prompt_adc}
\end{figure}

The energy deposition of the stopped muons in the HPGe detector is
$\rm 11.3\!\pm\!0.7\,MeV$ (see
Fig.~\ref{fig:muex_hpge_delayed_tdc_prompt_adc}).  The cut on the
energy of the prompt signal is $9.9 - 12.7$\,MeV.  The delayed signal
has to occur within the time interval of $\rm 3.4-7.2\,\mu{}s$ after
the prompt signal. The time distribution (see
Fig.~\ref{fig:muex_hpge_delayed_tdc_prompt_adc}) nicely shows the
expected exponential shape with $\rm \tau=2.21\pm0.02\,\mu{}s$.  For
shorter times the tail of the prompt signal still causes a varying
effective discriminator threshold thus the TDC spectrum deviates from
an exponential shape. The information from the NaI detector is used to
check the consistency of the analysis, but is not used for the
determination of the branching ratio.

After energy and time cuts $1.32\cdot10^6$ events remain.
Accounting for high energy positrons from muon decay causing a signal
in the veto counter S1, a 3\% correction results in $1.36\cdot10^6$
good muon decays for normalization.

\begin{figure}
  \centerline{\epsfig{file=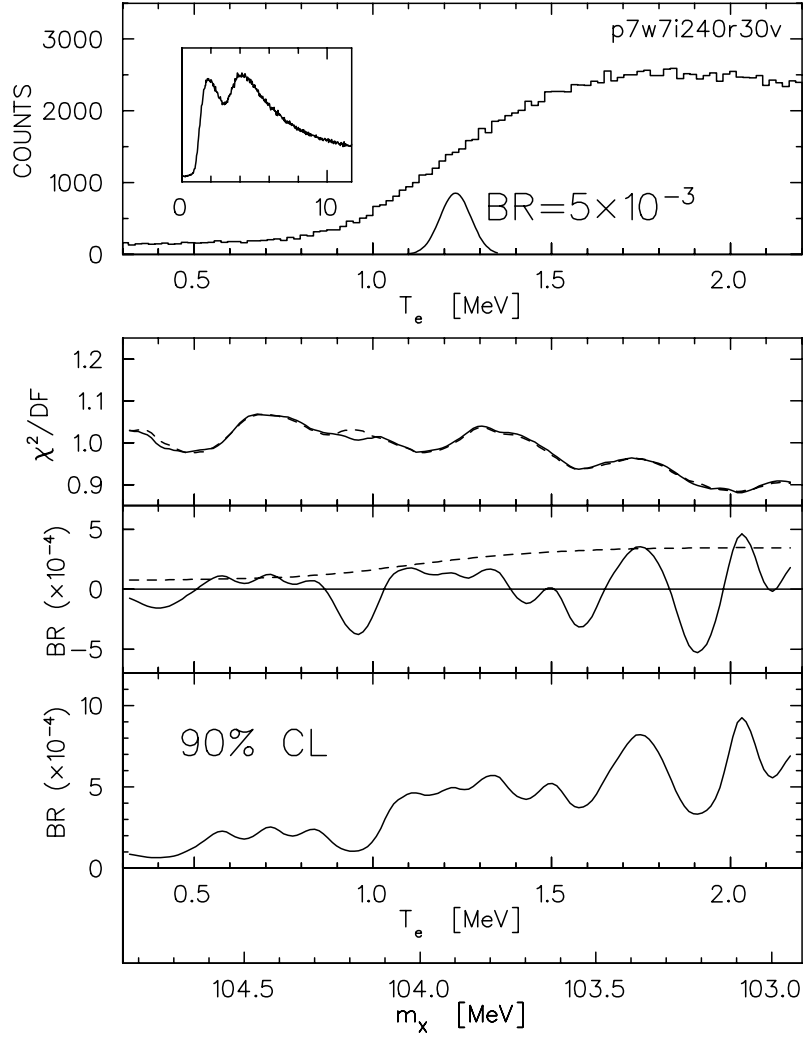,width=0.80 \textwidth}}
  \caption[]{Plots showing the energy deposition during the delayed
    gate in the HPGe detector (top) and fit results leading to upper
    limits for the branching ratio for the decay $\Pgmp\to\Pep\PX$.
    For the abscissa two corresponding scales, which are the same for
    all graphs (except for the inset at the top, which shows the full
    \Pep\ energy range recorded), are drawn, one is the positron
    kinetic energy $\rm T_\Pe$, the other the \PX\ boson mass $\rm
    m_\PX$. In the graph at the top the Gaussian centered at 1.23\,MeV
    gives the expected detector response if $\Pgmp\to\Pep\PX$ would
    contribute with a branching ratio of $5\cdot10^{-3}$.  The second
    graph shows the reduced $\chi^2$, dashed line for a
    polynomial-only fit, solid line for a combined polynomial and
    Gaussian fit. The third graph, with ordinate units already
    converted into branching ratio, shows the contents (solid line)
    and the error (dashed line) of the Gaussian from this fit.  The
    graph at the bottom gives the upper limit for a $\Pgmp\to\Pep\PX$
    decay branch at 90\% confidence level by applying the Bayesian
    method to the fit results.}
  \label{fig:muex_branching}
\end{figure}

GEANT \cite{geant:1998} based Monte Carlo studies have provided an
understanding of the shape of the delayed signal energy spectrum (see
inset in Fig.~\ref{fig:muex_branching}). The two peaks are due to an
asymmetric \Pgm\ stop distribution with respect to the symmetry plane
perpendicular to the beam axis of the cylindrical HPGe detector
resulting in different energy distributions for Michel positrons
emitted in the backward and forward hemispheres of the detector,
respectively.

The interaction of the annihilation \Pgg\ rays with the detector
has also been studied. For positrons in the considered energy range
the double escape probability is 40-44\% (no 511\,keV \Pgg\ rays
interacting in the HPGe), the single escape probability being a factor
4 lower. The search for $\Pgmp\to\Pep\PX$ events as described below
concentrates on double escape events.

Assuming a smooth and gently varying background as confirmed by the
Monte Carlo studies, the search for a peak structure in the delayed
signal energy spectrum (see Fig.~\ref{fig:muex_branching}) has been
done for energies from $0.3$ to $2.2$\,MeV. The lower energy limit is
given by the effective discriminator threshold, the upper energy limit
from the positron zero transmission range in germanium. Since the beam
muons are stopped after $\rm 2-3\,mm$ (2$\sigma$) in the HPGe
detector, and since the 2.2\,MeV electrons have a zero transmission
range of 2\,mm, this is the highest energy for which all positrons
remain within the detector volume thus completely depositing their
kinetic energy.

For all positron energies between 0.3 and 2.2\,MeV a typically
1.2\,MeV wide energy interval is chosen and a polynomial fitted to
this part of the spectrum.  For a polynomial of low order the fit has
an unrealistically high $\chi^2$.  Increasing the order of the
polynomial the resulting $\rm \chi^2/D.F.$ first decreases and then
remains roughly constant with values around one.

A polynomial of order seven was chosen as the lowest order to have a
suitable reduced $\chi^2$ (second graph in
Fig.~\ref{fig:muex_branching}).  Then a simultaneous fit of a Gaussian
(position and width fixed) and a polynomial provides the area and
error for a possible peak.  In the third graph of
Fig.~\ref{fig:muex_branching} these results have already been
converted in branching ratio (BR) units.  With a Bayesian
approach~\cite{barnett:1996} one can derive from these results an
upper limit with a given confidence level.  Shown on the bottom of
Fig.~\ref{fig:muex_branching} is the 90\% C.L.\ upper limit.

For the positron energy $\rm T_\Pe=1.23\,MeV$ corresponding to an \PX\ 
particle with mass $\rm m_\PX=103.9\,MeV/c^2$ as suggested by Gninenko
and Krasnikov~\cite{gninenko:1998} the 90\% C.L.\ upper limit for the
branching ratio in the decay $\Pgmp\to\Pep\PX$ is $\rm
BR=4.9\cdot10^{-4}$.


\section{Summary and Outlook}

Following the proposition that a new, weakly interacting boson \PX\ 
with mass $\rm m_\PX=103.9\,MeV/c^2$ produced in $\Pgmp\to\Pep\PX$
might be the reason for the observed anomaly in the KARMEN data, we
have searched for this two-body \Pgm\ decay branch by inspection of
the low energy end of the Michel spectrum. Utilizing a clean \Pgm\ 
beam from the $\mu{}E4$ channel at PSI and stopping the muons in a
planar HPGe detector this work is the first direct search for such an
exotic \Pgm\ decay process for \PX\ boson masses $\rm
103\,MeV/c^2<m_\PX<105\,MeV/c^2$ corresponding to positron energies
$\rm 0.3\,MeV<T_e<2.2\,MeV$.  Our first results give branching ratios
$\rm BR(\Pgmp\to\Pep\PX)<5.7\cdot10^{-4}$ (90\% C.L.) over most of the
accessible region, such excluding the simplest scenario for the \PX\ 
boson phenomenology suggested in Ref.~\cite{gninenko:1998}. By
refining the experimental method used in this experiment it will be
feasible to improve on this result.

\bigskip

\noindent We gratefully acknowledge valuable support from and
discussions with D.~Branford, M.~Daum, T.~Davinson, F.~Foroughi,
C.~Petitjean, D.~Renker, U.~Rohrer, and A.C.~Shotter. We also would
like to thank the Paul Scherrer Institut for assistance in setting up
this experiment in a very short time.


\end{document}